\documentclass[a4paper,10pt]{article}
\usepackage{graphicx}
\usepackage{comment}
\usepackage{amssymb}
\usepackage{amsmath}
\usepackage{nicefrac}
\usepackage{graphicx}
\usepackage{dcolumn}
\usepackage{bm}
\usepackage{comment}
\usepackage{multirow}
\usepackage{cite}

\begin{document}

\newcommand{\bin}[2]{\left(\begin{array}{c}\!#1\!\\\!#2\!\end{array}\right)}
\newcommand{\threej}[6]{\left(\begin{array}{ccc}#1 & #2 & #3 \\ #4 & #5 & #6 \end{array}\right)}
\newcommand{\sixj}[6]{\left\{\begin{array}{ccc}#1 & #2 & #3 \\ #4 & #5 & #6 \end{array}\right\}}
\newcommand{\regge}[9]{\left[\begin{array}{ccc}#1 & #2 & #3 \\ #4 & #5 & #6 \\ #7 & #8 & #9 \end{array}\right]}
\newcommand{\La}[6]{\left[\begin{array}{ccc}#1 & #2 & #3 \\ #4 & #5 & #6 \end{array}\right]}

\huge

\begin{center}
Some properties of Wigner $3j$ coefficients: non-trivial zeros and connections to hypergeometric functions\end{center}

\vspace{0.5cm}

\large

\begin{center}
Jean-Christophe Pain$^{a,b,}$\footnote{jean-christophe.pain@cea.fr}
\end{center}

\normalsize

\begin{center}
\it $^a$CEA, DAM, DIF, F-91297 Arpajon, France\\
\it $^b$Universit\'e Paris-Saclay, CEA, Laboratoire Mati\`ere sous Conditions Extr\^emes,\\
\it 91680 Bruy\`eres-le-Ch\^atel, France
\end{center}

\vspace{0.5cm}

\begin{abstract}
The contribution of Jacques Raynal to angular-momentum theory is highly valuable. In the present article, I intend to recall the main aspects of his work related to Wigner $3j$ symbols. It is well known that the latter can be expressed with a hypergeometric series. The polynomial zeros of the $3j$ coefficients were initially characterized by the number of terms of the series minus one, which is the degree of the coefficient. A detailed study of the zeros of the $3j$ coefficient with respect to the degree $n$ for $J=a+b+c\leq 240$ ($a$, $b$ and $c$ being the angular momenta in the first line of the $3j$ symbol) by Raynal revealed that most zeros of high degree had small magnetic quantum numbers. This led him to define the order $m$ to improve the classification of the zeros of the $3j$ coefficient. Raynal did a search for the polynomial zeros of degree 1 to 7 and found that the number of zeros of degree 1 and 2 are infinite, though the number of zeros of degree larger than 3 decreases very quickly as the degree increases. Based on Whipple's symmetries of hypergeometric $_3F_2$ functions with unit argument, Raynal generalized the Wigner $3j$ symbols to any arguments and pointed out that there are twelve sets of ten formulas (twelve sets of 120 generalized $3j$ symbols) which are equivalent in the usual case. In this paper, we also discuss other aspects of the zeros of $3j$ coefficients, such as the role of Diophantine equations and powerful numbers, or the alternative approach involving Labarthe patterns. 
\end{abstract}

\section{Introduction}\label{sec1}

As an atomic physicist, the theory of angular momentum has always been an important subject of research for me, for practical applications of course, but also for the elegance of the formalism, the surprises it holds and its unsolved mysteries. Jacques Raynal's works are among those which most caught my attention. I never met him, and a few years ago, I questioned colleagues from CEA Saclay to know more about him, but in vain. A few months ago, I learned by chance from Eric Bauge that Raynal had been a regular collaborator of the nuclear physics division of the CEA center where I work, in Bruy\`eres-le-Ch\^atel. Unfortunately Jacques died shortly after, before I could meet him.

In quantum mechanics, Clebsch-Gordan coefficients describe how individual angular momentum states may be coupled to yield the total angular momentum  state of a system \cite{EDMONDS57}. In the literature, Clebsch-Gordan coefficients are sometimes also known as Wigner coefficients or vector coupling coefficients. They are closely related to Wigner's $3j$ symbol \cite{BIEDENHARN81} by 

\begin{equation}\label{cle}
C_{a\alpha,~b\beta}^{c(-\gamma)}=(-1)^{a-b-\gamma}\sqrt{2c+1}\threej{a}{b}{c}{\alpha}{\beta}{\gamma},
\end{equation}

\noindent where $C_{a\alpha,~b\beta}^{c\gamma}$ is the Clebsch-Gordan coefficient and 

\begin{equation}\label{3jdb}
\threej{a}{b}{c}{\alpha}{\beta}{\gamma}
\end{equation} 

\noindent the $3j$ symbol for angular momenta $a$, $b$ and $c$ with respective projections $\alpha$, $\beta$ and $\gamma$. Explicit recursion formulae and closed-form factorial sum expressions for the Clebsch-Gordan coefficients have long been known \cite{LOUCK58}. Schulten and Gordon developed recursion relations suitable for the computer calculation of high-order $3j$ symbols, thus allowing sequences of coefficients to be calculated where one of the principal quantum numbers changes and the others remain fixed \cite{SCHULTEN76}. However, in this approach intermediate results need to be re-normalized to prevent the recurrence from diverging. Such problems were largely resolved by Luscombe and Luban who introduced a non-linear recurrence technique, although it is necessary in their approach to protect against possible division-by-zero problems \cite{LUSCOMBE98}. A practical method for calculating the Clebsch-Gordan coefficients in nested form for large arguments was published by Nasser et al. \cite{NASSER87}. In order to avoid the risk of numerical instabilities, several authors calculate Clebsch-Gordan coefficients using explicit factorial sum formulae, in which the factorials are expressed as products of prime numbers \cite{STONE80,LAI90,JOHANSSON16}. This allows many integer factors to be manipulated and ultimately cancelled symbolically. When coupled with a modern extended precision arithmetic library this also allows Clebsch-Gordan coefficients or $3j$ symbols with very large quantum numbers to be calculated accurately. However, this approach is computationally expensive, as only one Clebsch-Gordan coefficient or $3j$ symbol can be calculated at a time. Other authors have aimed to achieve a compromise between speed and accuracy by rewriting some of the factorial terms as binomial coefficients \cite{LAI90,GUSEINOV95,WEI99}, or as triangle coefficients in Regge squares \cite{ROOTHAAN93,TUZUN98,ROOTHAAN97} in order to allow common factors to be re-used. Ritchie \cite{RITCHIE11,RITCHIE19} presented simple two-term and three-term recursion formulae for calculating high order Clebsch-Gordan coefficients rapidly and accurately. More specifically, he expressed a Clebsch-Gordan coefficient as a sum of products of three binomial ones, called an un-normalized Clebsch-Gordan coefficient, the remaining factorial product terms being treated as a normalization factor. This separation of terms allows un-normalized Clebsch-Gordan coefficients to be calculated exactly in integer arithmetic using  simple three-term recursion formulae \cite{SURESH11}; the recursion is guaranteed to be stable for arbitrarily large quantum numbers. Such an approach follows a similar intuition to the recursive binomial sum method of Tuzun et al. \cite{TUZUN98} except that their approach requires several different formulae to initialize the recursion. Ritchie's method \cite{RITCHIE11,RITCHIE19} is particularly well suited to calculating lists of Clebsch-Gordan coefficients ``on-the-fly'' for different magnetic quantum numbers being projections of the same angular momentum, thus avoiding the need for large or complex in-memory storage schemes. Very recently, Xu proposed an improved recursive computation of Clebsch-Gordan coefficients \cite{XU20}; the method separates the recursion process into sign-recursion and exponent-recursion. The Clebsch-Gordan values can be obtained after the computation of their signs and exponents. The method, called ``sign-exponent'' recursion by the authors, removes the risk of generating numerical overflows and underflows, and seems in general more stable than the three-term linear recurrence method.  The Clebsch-Gordan coefficient itself is not involved in the recursion, except a non-trivial zero occurs. In the latter case, the sign value will become zero and the exponent $-\infty$, making the sign-exponent method meaningless. In that case, the original three-term linear recurrence relations of Schulten and Gordon \cite{SCHULTEN76} should be applied. 

For the $3j$ coefficients of SU(2), there exists a class of zeros which are called non-trivial, structural or polynomial zeros (due to their connection with Hahn polynomials \cite{KOORNWINDER81}), as opposed to trivial zeros due to a symmetry or a violation of one or more triangle relation. In the encyclopedia volume ``The Racah-Wigner Algebra in Quantum Theory'' \cite{BIEDENHARN81}, this subject appears as Special Topic 10. Zeros of $3j$ coefficients find many applications in atomic and nuclear physics \cite{JUDD70,DESHALIT63,GINOCCHIO80}. There is also a more fundamental mathematical motivation, since the non-trivial zeros of $3j$ coefficients correspond to roots of special polynomials. Moreover, non-trivial zeros have appeared in relation to exceptional Lie groups and algebra \cite{BIEDENHARN81,KOOZEKANANI74,VANDERJEUGT83,DEMEYER84,VANDENBERGUE84,VANDERJEUGT92}. Bowick \cite{BOWICK76} shortened the tables of zeros for $3j$ coefficients taking into account the Regge symmetries \cite{REGGE58,SHELEPIN64}. Rao and Rajeswari pointed out that the binomial form of the Clebsch-Gordan coefficient explicitely reveals further structural of trivial zeros, and that among the 39 reduced ``non-trivial'' zeros listed by Bowick \cite{BOWICK76} and Biedenharn and Louck \cite{BIEDENHARN81} for $0\leq J=a+b+c\leq 27$, 21 are trivial structural zeros \cite{RAO84}.

In a systematic approach, the non-trivial zeros of $3j$ coefficients can be classified by the minimum length of the single-sum expression for the coefficients. This minimum length corresponds to the number of terms when the coefficient is rearranged as a generalized hypergeometric series \cite{MAJUMDAR84,RAO85,RAO92b}. The number of terms in this sum minus one is called the degree or the weight of the coefficient \cite{LOUCK91}. Thus zeros of degree $n>0$ are by definition non-trivial zeros. The zeros of degree 1 of the $3j$ coefficient are quite easy to find, and explicit expression for them, not taking into account the Regge symmetries, have been provided by Varshalovich. Using the Pell equation \cite{MATTHEWS00}, Louck and Stein \cite{LOUCK87}, showed that there are infinite sequences of zeros of degree 2 for $3j$ coefficients.

Raynal investigated the distribution of the zeros of $3j$ coefficients with respect to their weight $n$ and the sum $J$ of the angular momentum quantum numbers ($J$ is invariant under Regge symmetries). Its first computer search for the zeros of the $3j$ coefficients with $J\leq 240$ gave many zeros of high degree. However, inspection showed that most of these can be written with small magnetic quantum numbers. This observation led him to postulate that, apart from the degree $n$ a new quantity, the order $m$ can be introduced to classify the zeros of $3j$ coefficients. In his terminology, $3j$ coefficients which never vanish are coefficients of order zero. He showed \cite{RAYNAL93} that recurrence relations can give zeros near the zeros of order 0 and defined the order $m$ of a zero. He also solved the equations for zeros of order 1 and gave them explicitly, and presented partial results for zeros of orders 2 and 3.

It is worth mentioning that Donley Jr. and Kim developed an alternative approach to the theory of Clebsch-Gordan coefficients in terms of combinatorics and convex geometry. They found new features including a censorship rule for zeros, a sequence of 36-pointed stars of zeros, and another proof of Dixon's identity. As a major application, the authors also reinterpreted the work of Raynal et al. on vanishing Clebsch-Gordan coefficients as a ``middle-out'' approach to computing a specific type of matrices \cite{DONLEY18a,DONLEY18b}. Recently, Donley pointed out that in the theory of Clebsch-Gordan coefficients, one may recognize the domain space as the set of weakly semi-magic squares of size three. He considered two partitions of this set: a triangle-hexagon model based on top lines and one based on the orbits under a finite group action, and gave a generating function that counts the so-called trivial zeros of Clebsch-Gordan coefficients \cite{DONLEY19}. 

The expression of a $3j$ coefficient in terms of a $_3F_2$ hypergeometric series is not unique. The same coefficient can be written by means of different $_3F_2$ series, all involving at least the same number of terms. Raynal \cite{RAYNAL78} has summarized all these expressions. One of the nine formulas of Raynal or one of the seven formulas of Varshalovich is of special interest for the search of zeros. It has previously been obtained by Bandzaitis and Yutsis \cite{YUTSIS65} and reads as

\begin{equation}
_3F_2\left[\begin{array}{c}
-n,-z,-J-1\\
-z-x,-z-t
\end{array};1
\right].
\end{equation}

\noindent When rewriting this as an alternating sum of integers as above, one can see that $J+1$ is a factor in all these integers, except the first. Therefore, if $J+1$ is a prime number, the sum cannot vanish. This result was obtained by Bryant and Jahn \cite{BRYANT60}. The conditions on the parameters of the $_{p+1}F_p(1)$ relate the problem of the zeros of degree 1 to the solutions of the homogeneous multiplicative Diophantine equations \cite{COHEN07,CHOUDRY10}, and the problem of zeros of degree 2 is related to the solutions of Pell's equation \cite{BEYER86}. The Diophantine approach shows that the number of zeros of the angular-momentum coefficients is denumerably infinite, and reveals a connection with powerful numbers in number theory. 

In Sec. \ref{sec2}, the relations between Clebsch-Gordan coefficients are outlined, as well as the link with expectation values. The properties and classification of weight-1 and weight-2 zeros are discussed in Sec. \ref{sec3}, as well as the connection with the Pell equation. An alternative way of finding the zeros, relying on the use of Labarthe patterns \cite{LABARTHE86,LABARTHE00,LAI05}, is presented in Sec. \ref{sec4}.

\section{Connection to hypergeometric series}\label{sec2}

\subsection{Role of symmetries}\label{sec21}

The $3j$ coefficient can be written as \cite{RAO93}:

\begin{eqnarray}\label{ser}
\threej{a}{b}{c}{\alpha}{\beta}{\gamma}&=&\delta_{\alpha+\beta+\gamma,0}(-1)^{a-b-\gamma}\Delta(abc)\nonumber\\
& &\times\left[(a+\alpha)!(b+\beta)!(c+\gamma)!\right]^{1/2}\nonumber\\
& &\times\left[(a-\alpha)!(b-\beta)!(c-\gamma)!\right]^{1/2}\nonumber\\
& &\sum_t(-1)^t\left[t!\prod_{k=1}^2(t-\eta_k)!\prod_{\ell=1}^3(\xi_{\ell}-t)!\right]^{-1}
\end{eqnarray}

\noindent where $\max(0,\eta_1,\eta_2)\leq t\leq\min(\xi_1,\xi_2,\xi_3)$ and 

\vspace{3mm}

(i) $\eta_1=a-c+\beta=(a-\alpha)-(c+\gamma)$,

\vspace{3mm}

(ii) $\eta_2=b-c-\alpha=(b+\beta)-(c-\gamma)$,

\vspace{3mm}

(iii) $\xi_1=a-\alpha, \xi_2=b+\beta, \xi_3=a+b-c$,

\vspace{3mm}

(iv) $\Delta(abc)=\left[\frac{(-a+b+c)!(a-b+c)!(a+b-c)!}{(a+b+c+1)!}\right]^{1/2}$.

\vspace{3mm}

\noindent The function $\Delta(abc)$ vanishes unless the triangle inequality is satisfied by the three angular momenta. The series part in Eq. (\ref{ser}) exhibits 12 symmetries, since it is invariant under permutation of the two $\eta$- parameters and the three $\xi$- parameters ($2!\times 3!=12$). However, these are not the ``classical'' symmetries of the $3j$ coefficient which are due to the invariance by the 3! column permutations and the mirror symmetry: $\alpha\rightarrow-\alpha$, $\beta\rightarrow-\beta$ and $\gamma\rightarrow-\gamma$. 
In 1958, Regge arranged the nine non-negative integer parameters, referred to by Racah \cite{RACAH42,VARSHALOVICH88}: $-a+b+c$, $a-b+c$, $a+b-c$, $a-\alpha$, $b-\beta$, $c-\gamma$, $a+\alpha$, $b+\beta$, $c+\gamma$ into a $3\times 3$ square symbol (see Refs. \cite{REGGE58,SHELEPIN64} and \cite{VARSHALOVICH88} $\S~8.3.1$ p. 237):

\begin{equation}\label{regge}
\regge{(-a+b+c)}{(a-b+c)}{(a+b-c)}{a-\alpha}{b-\beta}{c-\gamma}{a+\alpha}{b+\beta}{c+\gamma}=[R_{ik}]
\end{equation}

\noindent and noted that all the sums of columns and rows add to $J=a+b+c$ (a property of magic squares). The expression of the $3j$ in terms of Regge's $R_{ij}$ coefficients sn given in Ref. \cite{VARSHALOVICH88} p. 242 and in Ref. \cite{YUTSIS65} (6 different expressions). The Regge symbol has 72 symmetries (see p. 244 of Ref. \cite{VARSHALOVICH88}), being invariant to 3! column permutations, 3! row permutations and a reflection about the diagonal; in other words:

\begin{itemize}

\item Permutation of columns or rows: the coefficient remains unchanged for cyclic permutations, and is multipled by $(-1)^J$ ($J=\sum_kR_{ik}=\sum_iR_{ik}$) under non-cyclic permutations.

\item Transposition: the coefficient remains unchanged.

\end{itemize}

\noindent This yields $6\times 6\times 2=72$ coefficients.

The classical symmetries arise due to the 3! column permutations and the exchange of rows 2 and 3 in $[R_{ik}]$. Permuting the pairs $(a,\alpha)$, $(b,\beta)$ and $(c,\gamma)$ in Eq. (\ref{ser}) yields six series representations. They can be expressed in terms of the $[R_{ik}]$ as

\begin{eqnarray}\label{ser2}
\threej{a}{b}{c}{\alpha}{\beta}{\gamma}&=&\delta_{\alpha+\beta+\gamma,0}\prod_{i,k=1}^3\left[\frac{R_{ik}!}{(J+1)!}\right]^{1/2}(-1)^{\sigma(pqr)}\nonumber\\
& &\sum_s(-1)^s\left[s!(R_{2p}-s)!(R_{3q}-s)!(R_{1r}-s)!\right.\nonumber\\
& &\left.\times(s+R_{3r}-R_{2p})!(s+R_{2r}-R_{3q})!\right]^{-1}
\end{eqnarray}

\noindent for all six permutations of $(pqr)=(123)$ with

\begin{equation}
\sigma(pqr)=\left\{\begin{array}{l}
R_{3p}-R_{2q} \;\;\;\;\;\;\;\;\;\;\mathrm{for~ even~ permutations}\\
R_{3p}-R_{2q}+J \;\;\;\;\mathrm{for~ odd~ permutations}.\\
\end{array}\right.
\end{equation}

The 6 column permutations are in one-to-one correspondence with the 6 series representations. Each series representation exhibits 12 of the 72 different symmetries of the $3j$ coefficient. The 12 symmetries exhibited by any one of the six series representations arise due to

\begin{itemize}

\item (i) the combined operation of an odd column permutation and the mirror symmetry;

\item (ii) a Regge symmetry deduced from the nine relations $R_{\ell p}+R_{mp}=R_{nq}+R_{nr}$ for cyclic permutations of both $(\ell mn)$ and $(pqr)=(123)$, or a Regge symmetry on which is superposed a combined even column permutation and the mirror symmetry. 

\end{itemize}

As pointed out by Srinivasa Rao and Rajeswari \cite{RAO93}, there has been no discussion about the symmetries exhibited by the $_3F_2(1)$ forms, especially the van der Waerden form \cite{WAERDEN32} and the 12 classical symmetries of the $3j$ coefficient. There exist no explicit correspondence between these. The symmetries of the van der Waerden $_3F_2(1)$ form clearly reveal the symmetries discovered later by Regge. Let us replace the factorials in Eq. (\ref{ser}) by Gamma functions $\Gamma(z)$. If the argument $z$ is negative one uses the replacement 

\begin{equation}
\Gamma(1-z-n)=(-1)^n\frac{\Gamma(z)\Gamma(1-z)}{\Gamma(z+n)}
\end{equation}

\noindent and obtain, for permutations $(pqr)=(123)$ the set of six $_3F_2(1)$:

\begin{eqnarray}\label{ser3}
\threej{a}{b}{c}{\alpha}{\beta}{\gamma}&=&\delta_{\alpha+\beta+\gamma,0}\prod_{i,k=1}^3\left[\frac{R_{ik}!}{(J+1)!}\right]^{1/2}(-1)^{\sigma(pqr)}\nonumber\\
& &\times\frac{1}{\Gamma(1-\mathcal{A},1-\mathcal{B},1-\mathcal{C},\mathcal{D},\mathcal{E})}~_3F_2\left[\begin{array}{c}
\mathcal{A},\mathcal{B},\mathcal{C}\\
\mathcal{D},\mathcal{E}
\end{array};1
\right],\nonumber\\
\end{eqnarray}

\noindent where $\mathcal{A}=-R_{2p}$, $\mathcal{B}=-R_{3q}$, $\mathcal{C}=-R_{1r}$, $\mathcal{D}=1+R_{3r}-R_{2p}$ and $\mathcal{E}=1+R_{2r}-R_{3q}$. The latter form is equivalent to the van der Waerden form \cite{WAERDEN32} derived within the theory of invariants. In addition to this form, three other forms have been available in the literature, due respectively to Wigner \cite{WIGNER31}:

\begin{equation}
_3F_2\left[\begin{array}{c}
a-\alpha+1,-c-\gamma,a-b-c\\
-b-c-\alpha,a-b-\gamma+1
\end{array};1
\right],
\end{equation}

\noindent to Racah \cite{RACAH42}:

\begin{equation}
_3F_2\left[\begin{array}{c}
a+\alpha+1,-c+\gamma,-a+\alpha\\
-b-c+\alpha,b-c+\alpha+1
\end{array};1
\right]
\end{equation}

\noindent and to Majumdar \cite{MAJUMDAR58}:

\begin{equation}
_3F_2\left[\begin{array}{c}
a+b-c+1,-c-\gamma,a-b-c\\
-2c,a-c-\beta+1
\end{array};1
\right].
\end{equation}

The symmetries of the $3j$ coefficient were discussed in terms of a set of five canonical parameters by Lockwood \cite{LOCKWOOD76}. A detailed discussion regarding the one-to-one correspondence between the 72 symmetries of the $3j$ coefficient and the permutation of the numerator and denominator parameters of the set of six $_3F_2(1)$ can be found in the papers of Venkatesh \cite{VENKATESH78,VENKATESH80a,VENKATESH80b}. He has shown that the study of the symmetries of the $3j$ coefficient in terms of the set of six $_3F_2(1)$s derived by him introduces a six-to-one homomorphism of the 72-element group of symmetries of the $3j$ coefficient onto the 12 permutations of parameters of a single $_3F_2(1)$ series of the set. Holman and Biedenharn pointed out a deep connection between SU(1,1) and SU(2) unitary representations and between the corresponding Clebsch-Gordan coefficients \cite{HOLMAN66}. Such a connection can be visualized through analytic continuation in the representation parameters in such a way that discrete and continuous representations appear on the same footing. D'adda et al. \cite{DADDA74} studied the symmetry of a function which generalized the $3j$ coefficients of SU(1,1) \cite{UI70} and SU(2) involving discrete unitary representations. In their work on symmetries of extended $3j$ coefficients, a set of real variables is introduced to express the $3j$ coefficient of SU(2) in terms of entire functions proportional to $_3F_2(1)$. Husz\'ar pointed out the usefulness of the Thomae-Whipple functions \cite{BAILEY35} and showed that the symmetry group of order 72, discovered by by Regge, is a straightforward consequence of 6 forms of the 120 Thomae-Whipple functions. He found that if the Regge group is enlarged by the transformations $j\rightarrow -j-1$ (Yutsis' ``mirror'' symmetry \cite{YUTSIS65,BANDZAITIS64}), a group of order 1440 is obtained, which is exactly the group generated by the interrelations between the 120 Thomae-Whipple functions \cite{HUSZAR72}.

Raynal \cite{RAYNAL78} used the two-term relation for $_3F_2(1)$ given by Thomae (generalization of Dixon's theorem) and obtained in addition to the terminating $_3F_2(1)$ forms given above, four non-terminating $_3F_2(1)$ forms having all positive numerator parameters. He stated that these formulae could be used for negative quantum numbers and performed a systematic study of all possible formulae and the conditions for their validity, using Whipple's work \cite{WHIPPLE25} on the symmetries of the $_3F_2(1)$ functions. Whipple's parameters, according to Raynal, provide a better representation of the symmetry properties of the generalized $3j$ coefficient (where the angular momenta and their projections can take any complex value) than the Regge square symbol. This is due to the fact that Whipple's formalism includes Yutsis' ``mirror'' symmetry $j\rightarrow -j-1$ and indicates the breakdown of the usual rules when the usual relations between quantum numbers are not fulfilled. Raynal shows that to each $_3F_2(1)$, 12 generalized $3j$ coefficients can be associated by permutations of the three numerator and the two denominator parameters. Since Whipple found that there are 120 equivalent $_3F_2(1)$, Raynal argues that there are in all $120\times 12$ equivalent generalized $3j$ coefficients. He states that there is no need to consider only the $_3F_2(1)$ which are finite sums. It is worth mentioning, in that context, that Rashid \cite{RASHID86} used a transformation between a terminating $_3F_2(1)$ and a terminating Saalschutzian $_4F_3(1)$ to get summation-free (or closed-form) expressions for specific Clebsch-Gordan coefficients. 

\subsection{Expectation values}\label{sec22}

Several authors have noted that the expectation value of operator $r^k$, $r$ being the radial-position operator in spherical coordinates, for an hydrogenic atom (i.e. for Coulomb wave functions) can be written as a sum of products of three binomial factors \cite{BOCKASTEN74,BOCKASTEN76}. Others have written the expectation value as a $_3F_2$ with argument unity \cite{HEIM09}, analogous to $3j$ symbol (atomic units are used):

\begin{eqnarray}
\langle r^k\rangle&=&\frac{n^{k-1}}{2^{k+1}}\frac{(n+\ell+k+1)!}{(n+\ell)!}\nonumber\\
& &\times ~_3F_2\left[
\begin{array}{l}
-k-1,-k-1,\ell+1-n\\
1,-n-\ell-k-1\\
\end{array};1
\right].
\end{eqnarray} 

\noindent In 1973, Karassiov and Shelepin pointed out the interest of an intimate relation between the calculation of finite differences, hypergeometric series and the theory of Clebsch-Gordan coefficients \cite{KARASSIOV73}. In 1979, Varshalovich and Khersonskii found a simple connection between $\langle r^k\rangle=\langle n\ell|r^k|n\ell\rangle$ and $C_{\ell~n,(k+1)~0}^{\ell~n}$ \cite{VARSHALOVICH79}. In 2015, Varshalovich and Karpova found that both $C_{\ell'~n,(k+1)~0}^{\ell~n}$ and $\langle n\ell|r^k|n\ell'\rangle$ can be put in a form in which they are proportional to the hypergeometric function \cite{SLATER66,LUKE69}:

\begin{equation}
_3F_2\left(\begin{array}{l}
\ell+\ell'-k,\ell-\ell'-k-1,-k-1\\
n+\ell-k,-2k-2\\
\end{array};1\right).
\end{equation}

\noindent They obtained the following result \cite{VARSHALOVICH15}:

\begin{equation}
\frac{\langle n\ell|r^k|n\ell'\rangle}{C_{\ell'~n,(k+1)~0}^{\ell~n}}=\frac{i^{\Delta}}{2n\sqrt{2\ell+1}}\left(\frac{n}{2}\right)^kf_{\ell,\ell'}^k,
\end{equation}

\noindent where

\begin{equation}
f_{\ell,\ell'}^k=\left[\frac{(k+1-\Delta)!(k+1+\Delta)!(\ell+\ell'+k+2)!}{\left[(k+1)!\right]^2(\ell+\ell'-k-1)!}\right]^{1/2},
\end{equation}

\noindent with $\Delta=\ell-\ell'\geq$ 0, and the Clebsch-Gordan $C_{\ell'~n,(k+1)~0}^{\ell~n}$ is evaluated in a nonphysical region of its arguments, where the so-called triangle condition is not satisfied and the projections of angular momenta $\ell$ and $\ell'$ are larger than the angular momenta themselves. In this region, the Clebsch-Gordan coefficient vanishes, but $f_{\ell,\ell'}^k\times C_{\ell'~n,(k+1)~0}^{\ell~n}$ is non zero. For $(k+1)<0$, coefficient $C_{\ell'~n,(k+1)~0}^{\ell~n}$ transforms into $(-1)^{\ell-\ell'}C_{\ell'~n,-(k+2)~0}^{\ell~n}$ and $f_{\ell,\ell'}^k$ into $f_{\ell,\ell'}^{-k}$. For negative values, one has $(-a)!/(-b)!=(-1)^{b-a}(b+1)!/(a+1)!$.

All these features for the radial aspect of the hydrogen atom may seem surprising, given no obvious rotational symmetry considerations for the radial equation. However, they are understood through the recognition that the radial problem has the symmetry of the non-compact group O(2,1) \cite{HEIM09,JUDD70b,ARMSTRONG71}. This group's closed triplet of operators under commutation is very similar, apart from sign changes in the structure factors, to those of the angular momentum's O(3) triplet, which explains the appearance of such Clebsch-Gordan coefficients in radial matrix elements. It was shown that the relation between expectation values and Clebsch-Gordan coefficients holds also in the relativistic (Dirac) case, \cite{PAIN20}.

\section{Zeros of $3j$ symbols}\label{sec3}

\subsection{Weight-1 zeros}\label{sec31}

Brudno \cite{BRUDNO85a} has given two one-parameter formulae for the weight-1 polynomial zeros of the $3j$ coefficient:

\begin{equation}
\threej{3n}{2n+1}{n+1}{3n-1}{-2n}{1-n}
\end{equation}

\noindent and

\begin{equation}
\threej{2n+1}{2n}{2}{n+1}{-n}{-1}
\end{equation}

\noindent for $n\in\mathbb{N}$. His paper ends with the statement: ``Proof that these equations constitute the total solution to the linear nontrivial problem, however, has yet to be demonstrated''. Brudno defines the problem of finding non-trivial zeros to be ``linear'' when the sums for the calculation of the coefficients have two terms: the linear solutions are those, where both terms cancel each other. As pointed out by Lindner \cite{LINDNER85}, this problem has already been solved p. 39 of his book on angular momentum \cite{LINDNER84}. The zeros called linear solutions by Brudno exist if and only if the smallest element of the Regge symbol (see Eq. (\ref{regge})):

\begin{equation}
[R_{ik}]=\regge{R_{11}}{R_{12}}{R_{13}}{R_{21}}{R_{22}}{R_{23}}{R_{31}}{R_{32}}{R_{33}}
\end{equation}

\noindent is equal to 1 and the algebraic minor of that element is zero (e.g. if $R_{11}$=1 then $R_{22}R_{33}$ must be equal to $R_{23}R_{32}$ for a non-trivial zero). 

Let us set $x=b-\beta$, $y=c-\gamma$, $u=b+\beta$ and $v=c+\gamma$. Since the zeros of $3j$ coefficients may all be classified by their weight, which is defined as the smallest parameter in the Regge representation, finding the weight-1 zeros requires solving the equation \cite{BRUDNO85b}:

\begin{equation}
\regge{1}{(x+u-1)}{(y+v-1)}{(u+v-1)}{x}{y}{(x+y-1)}{u}{v}=0,
\end{equation}

\noindent which boils down to finding every set $(x,y,u,v)$ that satisfies the relation

\begin{equation}\label{vd}
(x+u)(y-v)=(x-u)(y+v).
\end{equation}

\noindent Applying the identity

\begin{equation}
4ab=(a+b)^2-(a-b)^2
\end{equation}

\noindent to both sides of Eq. (\ref{vd}) and rearranging terms,  we obtain the Diophantine equation $X^2+Y^2=U^2+V^2$, where $X=x+y+u-v$, $Y=x-y-u-v$, $U=x+y-u+v$ and $V=x-y+u+v$. The set of all solutions of the latter equation was given in 1906 by Pasternak in a form suitable for the present discussion \cite{DICKSON05}. Pasternak proved that all the solutions of $X^2+Y^2=U^2+V^2$ are given by

\begin{equation}
X=kr+\ell s, Y=\ell r-ks,\\
U=kr-\ell s, V=\ell r+ks,
\end{equation}

\noindent where $k, r, \ell, s$ are integers. Brudno and Louck established that this implies that all weight-1 zeros are of the form

\begin{equation}
x=\alpha\beta, y=\beta\delta, u=\alpha\gamma, v=\gamma\delta,
\end{equation}

\noindent where $\alpha$, $\beta$, $\gamma$ and $\delta$ assume all positive integer values \cite{BRUDNO85b}. This result can be established resorting to Bell's theorem \cite{BELL33}.

\subsection{Weight-2 zeros}\label{sec32}

\subsubsection{Pell diophantine equation and powerful numbers}

Following Louck and Stein \cite{LOUCK87}, it is convenient to use the variables $(u_1,u_2,x_1,x_2)$ given by the following Regge array:
 
\begin{equation}
\regge{2}{x_1}{(x_2+u_1-2)}{(u_1+u_2-2)}{x_2}{(x_1-u_2+2)}{(x_1+x_2-u_2)}{u_1}{u_2},
\end{equation}
 
\noindent where $x_1=a-b+c$, $x_2=b-\beta$, $u_1=b+\beta$ and $u_2=c+\gamma$. The domain of the variables $(u_1,u_2,x_1,x_2)\in \mathbb{N}^4$ is such that the remaining four entries in the Regge array also belong to $\mathbb{N}$. The polynomial of interest associated with the Regge array is given by
 
\begin{eqnarray}
Q_{u_1,u_2}(x_1,x_2)&=&2u_1(u_1-1)(x_1-u_2+1)(x_1-u_2+2)\nonumber\\
& &-4u_1u_2(x_1-u_2+2)x_2\nonumber\\
& &+2u_2(u_2-1)x_2(x_2-1).
\end{eqnarray} 
 
\noindent The zeros of the Diophantine equation $Q_{u_1,u_2}(x_1,x_2)=0$ give all nontrivial zeros for which 2 occurs in the Regge array, i.e. weight-2 zeros. If any of the entries in the Regge array is equal to one, the polynomial reduces to that for special weight-1 zeros. We restrict the determination of the zeros to $(u_1,u_2,x_1,x_2)\in \mathbb{N}^4$ but also $u_1\geq 2$, $u_2\geq 2$, $x_1\geq u_2$, $x_2\geq 2$. For completeness, one has

\begin{eqnarray}
\regge{2}{x_1}{(x_2+u_1-2)}{(u_1+u_2-2)}{x_2}{(x_1-u_2+2)}{(x_1+x_2-u_2)}{u_1}{u_2}&=&\threej{\frac{u_1+u_2}{2}}{\frac{x_1+x_2}{2}}{(x_1+x_2+u_1-u_2)}{\frac{u_1+u_2-4}{2}}{\frac{x_2-x_1}{2}}{\frac{(x_1-x_2-u_1-u_2+4)}{2}}\nonumber \\
&=&\frac{(-1)^{x_1+u_1+u_2}}{4}\left[\frac{2(u_1+u_2-2)!}{u_1!u_2!}\right]^{1/2}\nonumber \\
& &\times\left[\frac{(x_1-u_2+3)_{u_2-2}(x_2+1)_{u_1-2}}{(x_1+x_2-u_2+1)_{u_1+u_2+1}}\right]^{1/2}Q_{u_1,u_2}(x_1,x_2),\nonumber\\
& &
\end{eqnarray}

\noindent where $u_1\geq 2$, $u_2\geq 2$ and $(x)_a=x(x+1)\cdots(x+a-1)$ the usual Pochhammer symbol. $u_1$ and $u_2$ are regarded as parameters and $x_1$, $x_2$ as variables. After a change of variables and a few transformations, one gets that finding the weight-2 zeros boils down to the resolution of equation

\begin{equation}
x^2-Dy^2=N,
\end{equation}

\noindent where the integers $D$ and $N$ are given respectively by 

\begin{equation}
D=u_1u_2(u_1+u_2-1)
\end{equation}

\noindent and 

\begin{equation}
N=-u_1(u_1-1)^2(u_2-1)(u_1+u_2),
\end{equation}

\noindent where $(u_1,u_2)\in \mathbb{N}^2$ and $u_1\geq 2$, $u_2\geq 2$. Such an equation possesses two parametric solutions:

\begin{equation}
\begin{array}{ll}
x=u_1(u_1-1), & \;\;\;\;y=u_1-1,\\
x=u_1(u_1+2u_2-1), & \;\;\;\;y=u_1+1,
\end{array}
\end{equation}

\noindent and it can be proven by elementary methods that these two solutions are the only parametric solutions. In fact, each parametric solution must have $y=u_1+a$, $a\in\mathbb{Z}$, since this is the only way to cancel the term $-u_1^4u_2$ on the right-hand side identically by a term from the left-hand side.

Pell's equation is the Diophantine equation 

\begin{equation}
x^2-Dy^2=1. 
\end{equation}

\noindent The first treatment of this equation was actually given by Brouckner in 1657, but his solution was erroneously ascribed to Pell by Euler. Lagrange developed the theory of continued fractions, giving the actual method of finding the minimal solution, and published the first proof in 1766. However, the so-called cyclic method was already known to the Indian mathematicians Brahmagupta and Bhaskara in the twelfth century \cite{VANDERWAERDEN76}. Pell's equation has infinitely many solutions if $D$ is not a perfect square and none otherwise. This can be easily proven using Chebyshev polynomials. Chebyshev polynomials of the first kind are defined as

\begin{equation}
T_n(x)=\cos[n\arccos(x)],
\end{equation}

\noindent where $0\leq\arccos(x)\leq\pi$ and Chebyshev polynomials of the second kind as

\begin{equation}
U_{n-1}(x)=\frac{1}{n}T_n'(x)=\frac{\sin[n\arccos(x)]}{\sin[\arccos(x)]}.
\end{equation}

\noindent In the following we use Schur's notation $\mathcal{U}_n(x)=U_{n-1}(x)$. It is a fact that \cite{HUA82}, under our assumption, there always exists a solution to the Pell equation, and hence a solution for which $x$ is least, say $\left(x_0,y_0\right)$. Note that $x_0>1$. Consider the identity

\begin{equation}
T_n^2(x)-\left(x^2-1\right)\mathcal{U}_n^2(x)=1;
\end{equation}

\noindent then

\begin{equation}
1=T_n^2\left(x_0\right)-\frac{\left(x_0^2-1\right)}{y_0^2}\left[y_0^2~\mathcal{U}_n^2\left(x_0\right)\right].
\end{equation}

\noindent Since $\left(x_0^2-1\right)/y_0^2=D$ by definition, we conclude that for each $n\in\mathbb{N}$, the pair $\left\{T_n\left(x_0\right),y_0~\mathcal{U}_n\left(x_0\right)\right\}$ is a solution of Pell's equation which implies that the latter has an infinity of solutions. It is not difficult to show that all positive solutions of Pell's equation (for a non-square $D$) are given by $\left\{T_n\left(x_0\right),y_0~\mathcal{U}_n\left(x_0\right)\right\}$, $n\in\mathbb{N}$. If $(x_0,y_0)$ denotes its minimal solution (called also the fundamental solution) different from $(1,0)$, then all solutions $\left(x_n,y_n\right)$ are obtained by means of the recurrence relations

\begin{equation}
\begin{array}{l}
x_{n+1}=x_0x_n+Dy_0y_n,\;\;\;\;y_{n+1}=y_0x_n+x_0y_n\\
x_1=x_0,\;\;\;\;y_1=y_0.
\end{array}
\end{equation}

\noindent The main point is thus finding the fundamental solution, which can be achieved using continued fractions \cite{BOJU07}. According to Lagrange, the continued fraction of a quadratic irrational number (i.e. a real number satisfying a quadratic equation with rational coefficients but that is not rational itself) is periodic after some point; in our case, even more can be obtained, namely $\sqrt{D}=\left[a_0,\overline{a_1,\cdots,a_m,2a_0}\right]$. The numbers $a_j$ can be calculated recursively as follows. Let us set $a_0=\lfloor D\rfloor$, where $\lfloor x\rfloor$ stands for the integer part of $x$, and construct the sequences:

\begin{equation}
\begin{array}{ll}
P_0=0, & \;\;\;\; Q_0=1,\\
P_1=a_0, & \;\;\;\; Q_1=D-a_0^2,\\
P_n=a_{n-1}Q_{n-1}-P_{n-1}, & \;\;\;\; Q_n=\frac{D-P_n^2}{Q_{n-1}}.
\end{array}
\end{equation}

\noindent Then $a_n=\lfloor\frac{a_0+P_n}{Q_n}\rfloor$. Considering the further sequences:

\begin{equation}
\begin{array}{ll}
p_0=a_0, & \;\;\;\; q_0=1,\\
p_1=a_0a_1+1, & \;\;\;\; q_1=a_1,\\
p_n=a_np_{n-1}+p_{n-2}, & \;\;\;\; q_n=a_nq_{n-1}+q_{n-2},
\end{array}
\end{equation}

\noindent we get the identities:

\begin{equation}
p_n^2-Dq_n^2=(-1)^{n+1}Q_{n+1},
\end{equation}

\noindent which enables one to build solutions to the Pell-like equations. If $a_{m+1}=2a_0$, as happens for a quadratic irrational, then $Q_{m+1}=1$. In particular, we find that the fundamental solution is determined by the parity of $m$, as follows: $x_0=p_m$, $y_0=q_m$ if $m$ is odd and $x_0=p_{2m+1}$, $y_0=q_{2m+1}$ if $m$ is even. One has also $p_r/q_r=\left[a_0,a_1,\cdots,a_r\right]$. The general solution is obtained by means of the trick $x^2-Dy^2=\left(x_0^2-Dy_0^2\right)^n$ and setting

\begin{equation}
\begin{array}{l}
x+\sqrt{D}y=(x_0+\sqrt{D}y_0)^n\\
x-\sqrt{D}y=(x_0-\sqrt{D}y_0)^n,
\end{array}
\end{equation}

\noindent we obtain the family of solutions

\begin{equation}\label{sol}
\begin{array}{l}
x_n=\frac{(x_0+\sqrt{D}y_0)^n+(x_0-\sqrt{D}y_0)^n}{2}\\
y_n=\frac{(x_0+\sqrt{D}y_0)^n-(x_0-\sqrt{D}y_0)^n}{2}.
\end{array}
\end{equation}

The so-called negative Pell equation $x^2-Dy^2=-1$ can also be solved, but it does not always have solutions. A necessary condition for this equation to be solvable is that all prime factors of $D$ be of the form $4k+1$, and that $D$ $\not\equiv$ 0 mod 4. However, these conditions do not suffice, as can be seen from the equation for $D$=34, which has no solution. The method of continued fractions applies to these equations as well. In fact, for this Pell-type equation, we have the fundamental solution $x_0=p_m, y_0=q_m$ if $m$ is even. The equation has no solution if $m$ is odd, and the general solution is given again by (\ref{sol}). 

The Pell-like equation $x^2-Dy^2=N$ can be solved using the same ideas. If $|N|<\sqrt{D}$, then the equation has solutions if and only if

\begin{equation}
N\in\left\{Q_0,-Q_1,\cdots,(-1)^{m+1}Q_{m+1}\right\}.
\end{equation}

\noindent Moreover, if $N>\sqrt{D}$, the procedure is significantly more complicated \cite{DICKSON05}. It is clear that using the solutions $(x_n,y_n)$ to the companion Pell equation $x^2-Dy^2=1$ and a particular solution $(z,t)$ of the general Pell-type equation from above, we are able to find infinitely many solutions. Indeed, using the Brahmagupta identity which enables to us write that if $(p,q)$ is a solution of 

\begin{equation}
x^2-Dy^2=N
\end{equation}

\noindent and if $(r,s)$ is a solution of $x^2-Dy^2=1$, we have

\begin{equation}
(p^2-Dq^2)(r^2-Ds^2)=(pr\pm Dqs)^2-D(ps\pm qr)^2=N,
\end{equation}

 \noindent and the latter solutions by are obtained by means of the recurrence

\begin{equation}
\begin{array}{l}
z_n=x_nz\pm Dy_nt\\
t_n=x_nt\pm y_nz.
\end{array}
\end{equation}

\noindent However, even if we start with the minimal solution, we cannot always obtain all solutions of the equation using this recurrence. The reason is that we might well have several fundamental solutions. For instance, if $D$=10 and $N$=9, then we have the fundamental solutions $(7,2)$, $(13,4)$ and $(57,18)$. 

Golomb \cite{GOLOMB70} defined a powerful number to be a positive integer $r$ such that $p^2$ divides $r$ whatever the prime $p$ divides $r$, and discussed consecutive pairs of powerful numbers which fall into one of two categories; type 1: pairs of consecutive powerful numbers one of which is a perfect square, and type 2: pairs of consecutive numbers neither of which is a perfect square. He showed that there is an infinity of cases of type 1 by applying theory of the Pell equation. Walker \cite{WALKER76} gave all cases of type 1, and formulated all pairs of consecutive powerful numbers of type 2, through certain solutions of another Diophantine equation. This is a curiosity, which does not suggest any physical application yet, but to our knowledge the connection between polynomial zeros of Clebsch-Gordan coefficients and powerful numbers was never mentioned before. 
 
\subsection{Raynal's classification and the order of a zero}

When $a+b+c$ is odd, one has:

\begin{equation}
\threej{a}{b}{c}{0}{0}{0}=0.
\end{equation}

\noindent On the contrary, if $J=a+b+c$ is even, it can not vanish and has the value:

\begin{eqnarray}\label{meq01}
\threej{a}{b}{c}{0}{0}{0}&=&(-1)^{\frac{J}{2}}\left[\frac{(J-2a)!(J-2b)!(J-2c)!}{(J+1)!}\right]^{1/2}\nonumber\\
& &\times\frac{\left(\frac{J}{2}\right)!}{\left(\frac{J}{2}-a\right)!\left(\frac{J}{2}-b\right)!\left(\frac{J}{2}-c\right)!},
\end{eqnarray}

\noindent a result which can be deduced from Dixon's theorem for $_3F_2(1)$ \cite{DIXON02}. From the recurrence relations for the $3j$ coefficient, Raynal \cite{RAYNAL78} deduced three more sets of $3j$ which never vanish and these are

\begin{eqnarray}\label{meq02}
\threej{a}{b}{c}{-\frac{1}{2}}{0}{\frac{1}{2}}&=&(-1)^{\frac{J}{2}+1}\left[\frac{(J-2a)!(J-2b)!(J-2c)!}{(J+1)!(2a+1)(2c+1)}\right]^{1/2}\nonumber\\
& &\times\frac{2\left(\frac{J}{2}\right)!}{\left(\frac{J}{2}-a-\frac{1}{2}\right)!\left(\frac{J}{2}-b\right)!\left(\frac{J}{2}-c-\frac{1}{2}\right)!}\nonumber\\
& &
\end{eqnarray}

\noindent for even $J$ and for odd $J$:

\begin{eqnarray}\label{meq03}
\threej{a}{b}{c}{-\frac{1}{2}}{0}{\frac{1}{2}}&=&(-1)^{\frac{J}{2}+\frac{3}{2}}\left[\frac{(J-2a)!(J-2b)!(J-2c)!}{(J+1)!(2a+1)(2c+1)}\right]^{1/2}\nonumber\\
& &\times\frac{2\left(\frac{J}{2}+\frac{1}{2}\right)!}{\left(\frac{J}{2}-a\right)!\left(\frac{J}{2}-b-\frac{1}{2}\right)!\left(\frac{J}{2}-c\right)!}\nonumber\\
& & 
\end{eqnarray}

\noindent as well as

\begin{eqnarray}\label{meq04}
\threej{a}{b}{c}{0}{1}{-1}&=&(-1)^{\frac{J}{2}+\frac{1}{2}}\left[\frac{(J-2a)!(J-2b)!(J-2c)!}{(J+1)!\bar{b}\bar{c}}\right]^{1/2}\nonumber\\
& &\times\frac{2\left(\frac{J}{2}+\frac{1}{2}\right)!}{\left(\frac{J}{2}-a-\frac{1}{2}\right)!\left(\frac{J}{2}-b-\frac{1}{2}\right)!\left(\frac{J}{2}-c-\frac{1}{2}\right)!}.\nonumber\\
& & 
\end{eqnarray}

\noindent These non-zero $3j$ coefficients will be called zeros of order 0. Contiguous $3j$ coefficients satisfy the recurrence relations \cite{RAYNAL79}:

\begin{eqnarray}\label{recuray}
& &-S(a,b,c,\alpha,\beta,\gamma)\threej{a}{b}{c}{\alpha}{\beta}{\gamma}-T(a,b,\alpha,\beta)\threej{a}{b}{c}{\alpha-1}{\beta+1}{\gamma}\nonumber\\
&=&S(a,b,c,-\alpha,-\beta,\gamma)\threej{a}{b}{c}{\alpha}{\beta}{\gamma}+T(a,b,-\alpha,-\beta)\threej{a}{b}{c}{\alpha+1}{\beta-1}{\gamma}\nonumber\\
&=&-S(b,c,a,\beta,\gamma,\alpha)\threej{a}{b}{c}{\alpha}{\beta}{\gamma}-T(b,c,\beta,\gamma)\threej{a}{b}{c}{\alpha}{\beta-1}{\gamma+1}\nonumber\\
&=&S(b,c,a,-\beta,-\gamma,\alpha)\threej{a}{b}{c}{\alpha}{\beta}{\gamma}+T(b,c,-\beta,-\gamma)\threej{a}{b}{c}{\alpha}{\beta+1}{\gamma-1}\nonumber\\
&=&-S(c,a,b,\gamma,\alpha,\beta)\threej{a}{b}{c}{\alpha}{\beta}{\gamma}-T(c,a,\gamma,\alpha)\threej{a}{b}{c}{\alpha+1}{\beta}{\gamma-1}\nonumber\\
&=&S(c,a,b,-\gamma,-\alpha,\beta)\threej{a}{b}{c}{\alpha}{\beta}{\gamma}+T(c,a,-\gamma,-\alpha)\threej{a}{b}{c}{\alpha-1}{\beta}{\gamma+1},
\end{eqnarray}

\noindent where (we use the notation $\bar{x}=x(x+1)$):

\begin{equation}
S(a,b,c,\alpha,\beta,\gamma)=\frac{1}{2}\left(\bar{a}+\bar{b}-\bar{c}\right)+\alpha\beta+\frac{1}{3}(\alpha-\beta),
\end{equation}

\noindent and

\begin{equation}
T(a,b,\alpha,\beta)=\left[(a+\alpha)(a-\alpha+1)(b-\beta)(b+\beta+1)\right]^{1/2}.
\end{equation}

\noindent Using once the recurrence relation (\ref{recuray}), for given values of the triplet $(\alpha,\beta,\gamma)$, one gets six sets of $3j$ coefficients which can be expressed in terms of coefficients of order zero (Eqs. (\ref{meq01}, \ref{meq02}, \ref{meq03} and \ref{meq04})). They are the zeros of order 1.    

(i) Setting $\alpha=\beta=\gamma=0$ in the recurrence relations and using a symmetry for the $3j$ coefficient, we get for even $J$:

\begin{equation}
\threej{a}{b}{c}{0}{-1}{1}=\frac{\bar{a}-\bar{b}-\bar{c}}{2\left(\bar{b}\bar{c}\right)^{1/2}}\threej{a}{b}{c}{0}{0}{0}.
\end{equation}

(ii) Setting $(\alpha,\beta,\gamma)=(0,1,-1)$, in the recurrence relations and using a symmetry for the $3j$ coefficient, we get for odd $J$:

\begin{equation}
\threej{a}{b}{c}{0}{2}{-2}=\frac{\bar{a}-\bar{b}-\bar{c}+2}{\left[(\bar{b}-2)(\bar{c}-2)\right]^{1/2}}\threej{a}{b}{c}{0}{1}{-1}
\end{equation}

\begin{equation}\label{17}
\threej{a}{b}{c}{1}{1}{-2}=\frac{(b-a)(a+b+1)}{\left[\bar{a}(\bar{c}-2)\right]^{1/2}}\threej{a}{b}{c}{0}{1}{-1}.
\end{equation}

\noindent For the latter equation, a zero can be found only if $a=b$ and it is a trivial zero since $J$ is odd. Setting 

\begin{equation}
(\alpha,\beta,\gamma)=(0,1/2,-1/2)
\end{equation}

\noindent and using symmetries and relabellings for $a$, $b$ and $c$, we get five new relations. They are, for $J$ even:

\begin{equation}
\threej{a}{b}{c}{0}{\frac{3}{2}}{-\frac{3}{2}}=\frac{\bar{a}-\bar{b}-\bar{c}-\left(b+\frac{1}{2}\right)\left(c+\frac{1}{2}\right)+\frac{1}{2}}{\left[\left(\bar{b}-\frac{3}{4}\right)\left(\bar{c}-\frac{3}{4}\right)\right]^{1/2}}\threej{a}{b}{c}{0}{\frac{1}{2}}{-\frac{1}{2}},
\end{equation}

\noindent for $J$ odd:

\begin{equation}
\threej{a}{b}{c}{0}{\frac{3}{2}}{-\frac{3}{2}}=\frac{\bar{a}-\bar{b}-\bar{c}+(b+\frac{1}{2})(c+\frac{1}{2})+\frac{1}{2}}{\left[\left(\bar{b}-\frac{3}{4}\right)\left(\bar{c}-\frac{3}{4}\right)\right]^{1/2}}\threej{a}{b}{c}{0}{\frac{1}{2}}{-\frac{1}{2}},
\end{equation}

\noindent for $J$ even:

\begin{equation}
\threej{a}{b}{c}{\frac{1}{2}}{1}{-\frac{3}{2}}=\frac{\left(a+\frac{1}{2}\right)(a+c+1)-\bar{b}}{\left[\bar{b}\left(\bar{c}-\frac{3}{4}\right)\right]^{1/2}}\threej{a}{b}{c}{\frac{1}{2}}{0}{-\frac{1}{2}},
\end{equation}

\noindent for $J$ odd:

\begin{equation}
\threej{a}{b}{c}{\frac{1}{2}}{1}{-\frac{3}{2}}=\frac{\left(a+\frac{1}{2}\right)(a-c)-\bar{b}}{\left[\bar{b}\left(\bar{c}-\frac{3}{4}\right)\right]^{1/2}}\threej{a}{b}{c}{\frac{1}{2}}{0}{-\frac{1}{2}}
\end{equation}

\noindent and for $J$ even or odd:

\begin{equation}
\threej{a}{b}{c}{\frac{1}{2}}{-1}{\frac{1}{2}}=-\frac{\left(c+\frac{1}{2}\right)+(-1)^J\left(a+\frac{1}{2}\right)}{\bar{b}^{1/2}}\threej{a}{b}{c}{\frac{1}{2}}{0}{-\frac{1}{2}}.
\end{equation}

\noindent Clearly, the six equations above in section (ii), except (\ref{17}), give the zeros of the $3j$ coefficient or recurrence order 1.

The $3j$ coefficients or order 2 are obtained with recurrence relations involving $3j$ coefficients of recursion order 0 and 1 for the two other members. More generally, the $3j$ coefficients of order $m$ are obtained with recurrence relations involving $3j$ coefficients of orders $m-2$ and $m-1$ for the two other members. In order to characterize the order $m$ in another way, let us perform, in this Regge symbol, a transformation bringing the two rows or two column with minimum absolute difference (the sum of the absolute values of the differences member by member) to the last two rows. One then has

\begin{itemize}

\item if $\alpha,\beta$ and $\gamma$ are all integers: $m=\max(|\alpha|,|\beta|,|\gamma|)$ if $J$ is even, $m=\max(|\alpha|,|\beta|,|\gamma|)-1$ if $J$ is odd;

\item if $\alpha,\beta$ and $\gamma$ are not all integers: $m=\lfloor\max(|\alpha|,|\beta|,|\gamma|)\rfloor$.

\end{itemize}

Note that this definition gives the order $m=-1$ for the trivial zeros. A complete classification of order 2 and 3 coefficients has been obtained and Raynal \cite{RAYNAL93} found that there are 12 types of zeros of order 2 and 17 types of zeros of order 3. The problem of zeros of order 1 has been completely solved and the zeros of order 2 and 3 classified. While the zeros of degree 1, 2, 3 and 4 were found to be infinite in number, it is not known whether the number of zeros of degree $n>4$ is finite or infinite. If the zeros are arranged with $n>m$, for increasing values of $m>4$, it appears that there are no zeros of high order!

\section{An alternative approach: the Labarthe patterns}\label{sec4}

The $3j$ symbol can be expressed in terms of so-called ``Labarthe patterns'' (or $L$-patterns) \cite{ROOTHAAN97,LABARTHE86,LABARTHE00,LAI05}. The primitive $L$-patterns are

\begin{equation}
\begin{array}{ll}
e_1=\La{0}{\frac{1}{2}}{\frac{1}{2}}{0}{\frac{1}{2}}{-\frac{1}{2}}, & e_2=\La{\frac{1}{2}}{0}{\frac{1}{2}}{-\frac{1}{2}}{0}{\frac{1}{2}}\\
\\
e_3=\La{\frac{1}{2}}{\frac{1}{2}}{0}{\frac{1}{2}}{-\frac{1}{2}}{0}, & e_4=\La{0}{\frac{1}{2}}{\frac{1}{2}}{0}{-\frac{1}{2}}{\frac{1}{2}}\\
\\
e_5=\La{\frac{1}{2}}{0}{\frac{1}{2}}{\frac{1}{2}}{0}{-\frac{1}{2}}, & e_6=\La{\frac{1}{2}}{\frac{1}{2}}{0}{-\frac{1}{2}}{\frac{1}{2}}{0}\\
\end{array};
\end{equation}

\noindent they satisfy the linear dependency relation:

\begin{equation}\label{rulesum}
e_1+e_2+e_3=e_4+e_5+e_6=\La{1}{1}{1}{0}{0}{0}.
\end{equation}

\noindent The Labarthe pattern associated to the $3j$ symbol (\ref{3jdb}) can be expressed as a linear combination of primitive $L$-patterns according to

\begin{equation}
\La{a}{b}{c}{\alpha}{\beta}{\gamma}=\sum_{\lambda=1}^6n_{\lambda}e_{\lambda},
\end{equation}

\noindent yielding the Labarthe formula for the $3j$ symbol:

\begin{equation}\label{mis}
\threej{a}{b}{c}{\alpha}{\beta}{\gamma}=Q\sum_{n_1}\sum_{n_2}\cdots\sum_{n_6}\frac{(-1)^p}{n_1!n_2!\cdots n_6!},
\end{equation}

\noindent where the normalization factor $Q$ is given by 

\begin{equation}
Q=\frac{1}{\sqrt{T_{abc}T_{abc,\alpha\beta\gamma}^-T_{abc,\alpha\beta\gamma}^+}},
\end{equation}

\noindent with

\begin{equation}
T_{abc}=\frac{(a+b+c+1)!}{(b+c-a)!(c+a-b)!(a+b-c)!}
\end{equation}

\begin{equation}
T_{abc,\alpha\beta\gamma}^-=\frac{1}{(a-\alpha)!(b-\beta)!(c-\gamma)!}
\end{equation}

\begin{equation}
T_{abc,\alpha\beta\gamma}^+=\frac{1}{(a+\alpha)!(b+\beta)!(c+\gamma)!}.
\end{equation}

\noindent The number of terms in the summation is the number of partition of an integer, which can be approximated by the Hardy-Ramanujan formula \cite{HARDY18}:

\begin{equation}
p(n)\approx\frac{1}{4n\sqrt{3}}e^{\pi\sqrt{\frac{2n}{3}}}.
\end{equation}

\noindent We obtain a set of six liner equations on the $n_k$'s:

\begin{equation}
\begin{array}{l}
n_1=a+b-c-n_5\\
n_2=b+\beta-n_5\\
n_3=a-\alpha-n_5\\
n_4=c-b+\alpha+n_5\\
n_6=c-a-\beta+n_5.\\
\end{array}
\end{equation}

\noindent Therefore, doing that, the six-index sum is reduced to a single sum and the summation over $n_5$ must be restricted by the conditions $\max(a-c+\beta,b-c-\alpha)\leq n_5$ and $n_5\leq \min(a+b-c,b+\beta,a-\alpha)$, which enables one to recover the Racah expression

\begin{eqnarray}
\threej{a}{b}{c}{\alpha}{\beta}{\gamma}&=&\frac{1}{Q}\sum_{n_5}\frac{(-1)^{a-b-\gamma+n_5}}{n_5!(a+b-c-n_5)!}\nonumber\\
& &\times\frac{1}{(b+\beta-n_5)!(a-\alpha-n_5)!}\nonumber\\
& &\times\frac{1}{(c-b+\alpha+n_5)!}\frac{1}{(c-a-\beta+n_5)!}.\nonumber\\
& &
\end{eqnarray}

\noindent In order to find the polynomial zeros of a $3j$ coefficient, we are looking for the sextuplets $\{n_1, n_2, n_3, n_4, n_5, n_6\}$ such as

\begin{equation}
\La{(x+u)/2}{(y+v)/2}{(x+y+u+v-2)/2}{(x-u)/2}{(y-v)/2}{(-x-y+u+v)/2}=\sum_{\lambda=1}^6n_{\lambda}e_{\lambda}.
\end{equation}

\noindent We find the decomposition 

\begin{equation}
ve_1+xe_2+e_3+(y-1)e_4+(u-1)e_5
\end{equation}

\noindent and using $e_3=e_4+e_5+e_6-e_1-e_2$ (see Eq. (\ref{rulesum})), we get

\begin{equation}
(v-1)e_1+(x-1)e_2+ye_4+ue_5,
\end{equation}

\noindent and applying

\begin{equation}
\sum_{n_1}\sum_{n_2}\cdots\sum_{n_6}\frac{(-1)^p}{n_1!n_2!\cdots n_6!}=0
\end{equation}

\noindent with

\begin{equation}
n=\sum_{\lambda=1}^6n_{\lambda}\;\;\;\mathrm{and}\;\;\;p=\sum_{\lambda=4}^6n_{\lambda},
\end{equation}

\noindent we obtain the Diophantine equation $vx=uy$, which is equivalent to $(x+u)(y-v)=(x-u)(y+v)$. Using that representation, the zeros of $3j$ coefficients can be parametrized\footnote{In Refs. \cite{LAI90,ROOTHAAN97}, the authors have a factor $(n+1)!$ in the numerator of the argument of the sum in Eq. (\ref{mis}). Such a factor was included in the normalization coefficient by Labarthe. Both expressions are actually equivalent.}.

The problem of multiplicative Diophantine equations was related to the idea of reciprocal arrays by Bell \cite{BELL33}. Bell categorized multiplicative Diophantine equations into seven types and obtained the solutions for them in terms of the minimum number of necessary and sufficient parameters. This intertwining hampered Bell from providing a general induction proof for his main theorem regarding the number of parameters obeying certain greatest common divisor conditions for obtaining the complete set of solutions for the homogeneous multiplicative Diophantine equation of degree $n$, namely 

\begin{equation}
x_1x_2\cdots x_n= u_1u_2\cdots u_n    
\end{equation}

\noindent for $n>2$. This problem was solved by Rao et al. who reformulated the main theorem of Bell and provided a general induction proof for it by avoiding the use of reciprocal arrays \cite{RAO92}. This fundamental number theoretic result was then used to provide a complete solution to the problem of polynomial zeros of degree 1 of the $3n-j$ coefficients of quantum theory of angular momentum (see Ref. \cite{RAO93}, Chapter 6).

\section{Conclusion}\label{sec5}

The contribution of Jacques Raynal to the theory of angular momentum is highly valuable. For instance, he performed a detailed study of the zeros of the $3j$ coefficient with respect the the degree $n$ for $J=a+b+c\leq 240$ ($a$, $b$ and $c$ being the angular momenta in the first line of the $3j$ symbol) \cite{RAYNAL93} and noticed that most zeros of high degree had small magnetic quantum numbers. This led him to define the order $m$ to classify the zeros of the $3j$ coefficient. He did a search for the polynomial zeros of degree 1 to 7 and found that the number of zeros of degree 1 and 2 are infinite, though the number of zeros of degree larger than 3 decreases very quickly. Based on Whipple' symmetries of hypergeometric $_3F_2$ functions with unit argument, he generalized the Wigner $3j$ symbols to any arguments and pointed out that there are twelve sets of ten formulas (twelve sets of 120 generalized $3j$ symbols which are equivalent in the usual case). 

The contribution of Jacques Raynal to the field is not limited to $3j$ coefficients; he also brought major advances to the theory of $6j$ symbols, as well as to Mohinsky coefficients \cite{RAYNAL76}. In the theoretical approach of the three-body problem, he introduced, with Janos Revai, the transformation bracket from one set of Jacobi coordinates to another set of a three-body system for the hyperspherical harmonic functions is called Raynal-Revai coefficient \cite{RAYNAL70}. An austrian friend of Raynal, Heinrich von Geramb, told him that professors Clebsch and Gordan from G\"ottingen were surprised when they were informed by colleagues that some coefficients were named after them by the participants to a conference. The same thing happened to Jacques Raynal when Revai, back from a conference in Dubna in 1970, announced him that their transformation bracket was named ``Raynal-Revai'' coefficient \cite{RAYNAL_INT}...

\section*{Acknowledgements}

I would like to thank Eric Bauge, Val\'erie Lapoux and Nicolas Alamanos for their invitation to participate to the issue on the topic ``Nuclear Reaction Studies: a Tribute to Jacques Raynal''. I am also indebted to Jean-Jacques Labarthe for helpful discussions.

\appendix

\section{The Whipple parameters}

Whipple \cite{WHIPPLE25} introduced six parameters $r_i$ ($i=0,1,2,3,4,5$) such that

\begin{equation}\label{c1}
\sum_{i=0}^5r_i=0.
\end{equation}

\noindent Setting $\alpha_{\ell mn}=1/2+r_{\ell}+r_m+r_n$ and $\beta_{mn}=1+r_m-r_n$, he defined the function

\begin{equation}\label{fp}
F_p(\ell;mn)=\frac{1}{\Gamma(\alpha_{ijk},\beta_{m\ell},\beta_{n\ell})}~_3F_2\left[\begin{array}{c}
\alpha_{imn},\alpha_{jmn},\alpha_{kmn}\\
\beta_{m\ell},\beta_{n\ell}
\end{array};1
\right],
\end{equation}

\noindent where $i, j, k$ are used to represent those three numbers out of the six integers $0, 1, 2, 3, 4, 5$ not already represented by $\ell$, $m$ and $n$. By changing the signs of all the $r_i$ parameters and using the constraint (\ref{c1}), Whipple defined another function \cite{WHIPPLE25}:

\begin{equation}\label{fn}
F_n(\ell;mn)=\frac{1}{\Gamma\left(\alpha_{\ell mn},\beta_{\ell m}, \gamma_{\ell n}\right)}.
\end{equation}

\noindent By permutation of the suffixes $\ell$, $m$, $n$ over the six integers $0, 1, 2, 3, 4, 5$, sixty $F_p$ functions and sixty $F_n$ functions can be written down. If there is no negative integer in the numerator parameters, these series converge only if the real parts of $\alpha_{ijk}$ in (\ref{fp}) and $\alpha_{\ell mn}$ in (\ref{fn}) are positive.

\end{document}